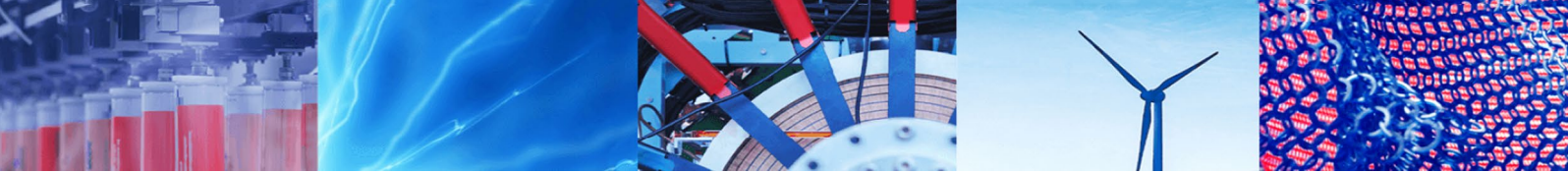

Research Article

# Future developments in standardisation of cyber risk in the Internet of Things (IoT)

Petar Radanliev[1] · David C. De Roure[1] · Jason R. C. Nurse[2] · Rafael Mantilla Montalvo[3] · Stacy Cannady[3] · Omar Santos[3] · La'Treall Maddox[3] · Peter Burnap[4] · Carsten Maple[5]

**Abstract**
In this research article, we explore the use of a design process for adapting existing cyber risk assessment standards to allow the calculation of economic impact from IoT cyber risk. The paper presents a new model that includes a design process with new risk assessment vectors, specific for IoT cyber risk. To design new risk assessment vectors for IoT, the study applied a range of methodologies, including literature review, empirical study and comparative study, followed by theoretical analysis and grounded theory. An epistemological framework emerges from applying the constructivist grounded theory methodology to draw on knowledge from existing cyber risk frameworks, models and methodologies. This framework presents the current gaps in cyber risk standards and policies, and defines the design principles of future cyber risk impact assessment. The core contribution of the article therefore, being the presentation of a new model for impact assessment of IoT cyber risk.

## 1 Introduction

There is a strong interest in industry and academia to standardise existing cyber risk assessment standards. Standardisation of cyber security frameworks, models and methodologies is an attempt to combine existing standards. This has not been done until present. Standardisation in this article refers to the compounding of knowledge to advance the efforts on integrating cyber risk standards and governance, and to offer a better understanding of cyber risk assessments. Here we combine literature analysis [1] with epistemological analysis, and an empirical [2] with a comparative study [3]. The empirical study is conducted with fifteen national high-technology (high-tech) strategies, seven cyber risk frameworks and two cyber risk models. The comparative study engages with fifteen high-tech national strategies. The epistemological analysis and an empirical study seek to probe the current understanding of cyber risk impact assessment.

To adapt the current cyber security standards, firstly the specific IoT cyber risk vectors need to be identified. By risk vectors, we refer to Internet of Things (IoT) attack vectors from particular approach used, to exploit big data vulnerabilities [4]. Subsequently, these specific risk vectors need to be integrated in a holistic cyber risk impact assessment model [5].

Documented process represents a new design for mapping and optimising IoT cyber security and assessing its associated impact. We discuss and expand on these further in the remainder of this article. The research article is structured in the following format. In Sect. 2 we present the research methodology. In Sect. 3 we conduct literature review. In Sect. 4 we propose the IoT cyber risk vectors by conducting a comparative study of national high-tech strategies and initiatives. In Sect. 5 we propose the design principles for impact assessment of IoT cyber risk by conducting empirical study cyber security frameworks, methods and quantitative models. In Sect. 6 we evaluate the design principles by conducting theoretical analysis to uncover the best method to define a unified cyber risk assessment. In Sect. 7 we propose a new epistemological framework for cyber risk assessment standardisation and we discuss the new impact assessment principles. In Sect. 8 we present the conclusions and limitations of the research.

✉ Petar Radanliev, petar.radanliev@oerc.ox.ac.uk | [1]Department of Engineering Sciences, Oxford e-Research Centre, University of Oxford, 7 Keble Road, Oxford OX1 3QG, UK. [2]School of Computing, University of Kent, Kent, UK. [3]Cisco Research Centre, Research Triangle Park, Durham, NC 27709, USA. [4]School of Computer Science and Informatics, Cardiff University, Cardiff, UK. [5]WMG Cyber Security Centre, University of Warwick, Coventry, UK.

## 2 Literature review

Literature review of academic and industry literature from several different countries is undertaken to advance the epistemological framework into a design model.

### 2.1 Recent literature on this subject

The increasing number of high-impact cyber-attacks has raised concerns of the economic impact [6–9] and the issues from quantifying cyber insurance [10]. This triggers questions on our ability to measure the impact of cyber risk [11–15]. The literature review is focused on defining the IoT risk vectors which are often overlooked by cyber security experts. The IoT risk vectors are investigated in the context of Social Internet of Things [16], the Industry 4.0 (I4.0) and the Industrial Internet of Things (IIoT). In the Social Internet of Things, the IoT is autonomously establishing social relationships with other objects, and a social network of objects and humans is created [17, 18]. The I4.0 is also known as the fourth industrial revolution and brings new operational risk for connected digital cyber networks [19]. Finally, the IIoT represents the use of IoT technologies in manufacturing [20].

#### 2.1.1 Cyber risk in shared infrastructure from autonomous IoT

The cyber risk challenges [21] from IoT technological concepts, mostly evolve around the design [22] and the potential economic impact (loss) from cyber-attacks [23, 24]. IoT revolves around machine-to-machine [25, 26] and cyber-physical systems (CPS) [27]. Similarly, the IoT is based on intelligent manufacturing equipment [28–32], creating systems of machines capable of interacting with the physical world [33]. The integration of such technologies creates new cyber risk, for example from integrating less secured systems [34]. Incorporating the cyber element in manufacturing, for instance, also bring an inherent cyber risk [35]. There are multiple attempts in literature where existing models are applied understand the economic impact of cyber risk [36]. However, these calculations largely ignore the cyber risk of sharing infrastructure [37], such as IoT infrastructure [11, 12, 38], [39–46]. Understanding the shared risk is vital for risk assessment [47], but the cyber risk estimated loss range can vary significantly [48]. Furthermore, there is no direct correlation between cyber ranking [49] and digital infrastructure [49], thus contradicting the argument that cyber risk is related to integrating new technologies [30]. It seems more likely that the cyber challenges are caused by the adoption and implementation [50] and the cost of smart manufacturing technologies [51].

#### 2.1.2 Cyber risk and IoT cloud technologies

If the Cloud architecture is properly engineered, the security of the cloud instance is adequately maintained and the connectivity from cloud to Thing can be assured, then cyber risks can be reduced with cloud technologies [52]. To ensure cyber risk is reduced, cloud technologies should be supported with: internet-based system, service platforms [53], processes, services [54], for machine decision making [55]. Creating cyber service architecture [56] and cloud distributed manufacturing planning [57]. Cyber risk mitigation also require compiling of data, processes, devices and systems [58]. IoT technologies need to be supported with a life cycle process for updating the list of assets that are added to the network across multiple time-scales [59–61]. IoT cyber risk is also present in components modified to enable a disruption [37, 62]. One option by which such risk could be mitigated is to consider the standardisation of the IoT design and process [47]; unfortunately however, such system security is complex [5] and risk assessing IoT systems is still a key problem in research [13]. Nevertheless, cyber networks need to be secure [63], vigilant [64], resilient [65] and fully integrated [66–68]. Therefore, the IoT need to encompass the security and privacy [69], along with electronic [70] and physical security of real-time data [71, 72].

The IoT consists of heterogeneous cloud technologies and varying lifecycle of the IoT devices, the question of value [73–75] in inheriting outdated data [76] where machines store knowledge and create a virtual living representation in the cloud [27]. The access to existing knowledge could be of value to design more resilient systems and processes in the future.

### 2.1.3 Cyber risk from social machines and real-time technologies

Cyber risk emerges from the Web [77], but also from any interface to a digital processing component, wired and wireless and the entire Web can be perceived as a social machine [78]. The term social machines in the context of this paper is used in relation to systems that depend on interaction between humans and technology and enable real time output or action, such as Facebook and Twitter. Social machines [66] are vulnerable to cyber risks, because of the connection between physical and human networks [65] operating as systems of systems [79] and mechanisms for real-time feedback [64]. Cyber risk from real-time IoT technology [80] requires information security for data in transit [57]. In addition, access control is required for granting or denying requests for information and processing services [81]. Despite expectations that information security and access control for social machines exists, the business of personal data has triggered many privacy concerns for social machines such as Facebook and Google [82]. Some of these concerns have already materialised [83, 84]. IoT brings inherent cyber risks which require appropriate cyber recovery plans. The relationship between IoT cyber risk assessment and recovery planning emerges from new processes, such as machine learning, that can be used to patch known vulnerabilities in real-time.

## 2.2 IoT cyber risk vectors from the literature review

The IoT cyber risk vectors relate to the overall aim of defining the design principles for cyber risk impact assessment. Prior to assessing the impact, we required an understanding of the IoT risk.

A list of IoT cyber risk vectors derive from the literature review.

- The cloud technologies enhance cyber security but amplify IoT cyber risk [31, 52, 66, 85, 86].
- IoT depends on real-time data, but real-time data amplifies IoT cyber risk [71, 72].
- IoT cyber risk mitigation needs autonomous cognition, but autonomous machine decisions amplify IoT cyber risk [28, 53, 87–89].
- These IoT cyber risk vectors are not clearly visible and focus should be on the communications risk; whether conventional wired (broadband or IP networks) or wireless (W-Fi, Bluetooth and 3G/4G/5G)—the connectivity is one of the weak spots [70].

While there are many more cyber risk vectors, analysing every single risk vector was considered beyond the scope of this study and the focus was placed on the most prominent vectors as identified in the literature. The idea was to identify a risk assessment process that can be applied by future researchers to many different risk vectors. The IoT risk vectors outlined above are analysed in the following section through comparative analysis of cyber risk in high-tech strategies.

## 3 Methodology

The methods applied in this study consist of literature review, comparative study, empirical analysis, theoretical and epistemological analysis and case study workshops. The selection of methodologies is based on their flexibility to be applied simultaneously to analyse the same research topic from different perspectives. We use practical studies of major projects in the I4.0 to showcase recent developments of IoT systems in the context of I4.0 high-tech strategies. We need practical studies to bridge the gaps, to assess the impact and overcome some of the cyber risk limitations and to construct the relationship between IoT and high-tech strategies. The proposed design principles support the process of building a holistic IoT cyber risk impact assessment model.

## 3.1 Theoretical analysis

The methodology applies theoretical analysis through logical discourse of knowledge, also known as epistemological analysis. An epistemological analysis enables an investigation on how existing knowledge is justified and what makes justified beliefs justified [90], what does it mean to say that we understand something [91] and how do we understand that we understand.

The methodology reported here has two objectives. The first objective is to enable an up-to-date overview of existing and emerging cyber risk vectors from IoT advancements, which includes cyber-physical systems, the industrial Internet of things, cloud computing and cognitive computing [53, 92, 93]. If we were performing a vector specific analysis of risk for the Internet of Things, we would include examining risk vectors related to consumer IoT and specific high-risk verticals like eHealth and Smart Cities. But this study is focused on the developing an economic impact assessment of IoT cyber risk as a component in the context of other emerging technologies. This methodological approach proposes a new design for assessing the impact of cyber risk and promotes the adaptation of existing cyber risk frameworks, models and methodologies. The second objective is to enable the adaptation of the best cyber security practices and standards to include cyber risk from IoT vectors.

The methodology begins with an academic and industry literature review on IoT cyber risk. A comparative study [3] classifies the cyber risk vectors, specific to the IoT, based on the current technological trends. An empirical study [2] categorises cyber risk frameworks, methodologies, systems, and models (particularly those that are quantitative). Afterwards, the compounded findings are compared with the existing standards through a grounded theory assessment method. This is followed by a theoretical analysis to uncover the best method to define a unified cyber risk assessment. The objective of the methodology is to synthesise and to build upon knowledge from existing cyber risk standards.

## 4 Comparative study on IoT cyber risk in high-tech strategies

This section represents a comparative study [3] of national high-tech strategies, because the IoT is strongly represented in the Industry 4.0. The selection of high-tech strategies—sources for analysis is based on the richness of the documented processes. The comparative study is applied on a range of IoT high-technology strategies to enhance the framework and to build upon previous literature on this subject [24]. Defining the most prominent IoT cyber risk vectors is of crucial importance to understanding IoT cyber risk, because IoT cyber risk is often invisible to cyber security experts. In this section, the study intent is to analyse Industry 4.0 and present it as an example of how risk assessment takes place at the national level.

### 4.1 Understanding IoT cyber risk in national high-tech strategies

The current direction of impact assessment from IoT cyber risk, seems to be decided by assessment activities, e.g. workgroups [94] or testbeds [20], supported by economic assessments [95]. In some strategies, impact is decided by assessing key projects in the digital industry, e.g. Fabbrica Intelligente [96] and Industrie 4.0 [97].

The different approaches to impact assessment, could be resulting from the differences in IoT focus. The Industrial Internet Consortium [20, 98] focuses on promoting core IoT industries; while the New France Industrial (NFI) [99], the High Value Manufacturing Catapult (HVM) [100] and the National Technology Initiative [101], all focus on promoting the development of key IoT technologies. Another high-tech strategy, Made in China 2025 [102], promotes tech sectors, while the Made Different [103] promotes key IoT transformations.

The diversity of the approaches can also be identified in the less evolved in identifying IoT cyber risk vectors (e.g. The Netherlands—Smart Industry [104]; Belgium—Made Different [103]; Spain—Industrie Conectada [92]; Italy—Fabbrica Intelligente [96]; G20—New Industrial Revolution [105]). This could be because some high-tech strategies lack documentation and appear disorganised. Such arguments are present in literature [106].

The Industrie 4.0 [86, 107]; the report by Department for Culture, Media and Sport (DCMS) [108], and the Industrial Value Chain Initiative (IVI) [94, 109] promote different risk vectors than the Russian National Technology Initiative (NTI) [101].

In some strategies, e.g. the Advanced Manufacturing Partnership [110], these differences are understandable, because one strategy would have evolved into a new high-the strategy, e.g. IIC [20], or are very narrowly focused on futuristic IoT technologies, e.g. New Robot Strategy [111]; Robot Revolution Initiative [112]; and the IoT technologies do not yet exist. Hence, we can only speculate on the expected cyber risks [11, 12].

The Table 1 summarises the analysis of the comparative study. The most prominent IoT cyber risk vectors derive from the analysis and are presented in a comparative decomposition approach. The aim of the comparative analysis and decomposition is to show the IoT cyber risk vectors and areas not covered (gaps) in national high-tech strategies. Secondly, the comparative analysis and decomposition enables visualising how the areas not covered in one high-tech strategy, have been addressed in other high-tech strategies. Therefore, the comparative study enables standardisation of approaches. The Table 1 enables policy makers to firstly identify the gaps and secondly to identify the best approach to address individual risk vectors. However, the analysis in Table 1 is limited to the most prominent vectors as identified in existing literature previously in Sect. 3.1.

To provide clarity on the areas not covered (gaps), the IoT cyber risk vectors are used as reference categories (Table 1), for decomposing the IoT cyber risk into sub-categories. The sub-categories are used for defining various IoT cyber risks vectors and for clarifying different and sometimes contrasting understanding of IoT cyber risks. The comparative study in Table 1, follows the grounded theory approach [113], and categorises the areas not covered in IoT risk vectors, to construct the cyber assessment design principles. In the following sections, a more general assessment is being presented and the national plans analysed are presented in a broader sense that take in more of the landscape of IoT implementation.

**Table 1** Analysis of IoT cyber risk vectors in high-tech national strategies

| | | | | |
|---|---|---|---|---|
| *Most prominent IoT cyber risk vectors* | | | | |
| Vectors | Vector 1 | Vector 2 | Vector 3 | Vector 4 |
| Risk vectors | Cloud | Real-time | Autonomous | Recovery |
| *IoT cyber risk vectors in documented and evolved high-tech strategies* | | | | |
| High-tech strategies | Vector 1 | Vector 2 | Vector 3 | Vector 4 |
| USA—(1) Industrial internet consortium [20] | Cloud-computing platforms | Operational models in real time; Customised products in real time | Fully connected and automated production line; Highly automated environments | Disaster recovery |
| (2) Advanced manufacturing partnership [110] | Not covered | Not covered | Not covered | NIST |
| UK—(1) UK digital strategy [108] | Cloud technology skills; Cloud computing technologies; Cloud data centres; Cloud-based software; Cloud-based computing; Cloud guidance | Digital real-time and interoperable records; Platform for real-time information | Robotics and autonomous systems; Support for robotics and artificial intelligence; Automation of industrial processes; Active cyber defence | Not covered |
| (2) Catapults [100] | Not covered | Not covered | Automation | Economic impact |
| Japan—(1) industrial value chain initiative [94] | Cloud enabled monitoring; Integration framework in cloud computing | Not covered | Factory automation; Robot program assets | Not covered |
| (2) New robot strategy (NRS) [111] | Not covered | Not covered | Robots innovation hub; Robot society; Robotics in IoT | Not covered |
| Robot revolution initiative (RRI) [112] | Society 5.0 | Connected industries | IoT in robotics | Not covered |
| Germany—Industrie 4.0 [97] | Cloud computing; cloud-based security networks | CPS systems | Automated production; Automated conservation of recourses | Not covered |
| Russia—National technology initiative (NTI) [101] | Not covered | Not covered | Artificial intelligence and control systems | Not covered |
| France—New france industrial (NFI) [99] | Not covered | Not covered | Automation and robotics | Impact assessment |
| *IoT cyber risk vectors in emerging and less evolved high-tech strategies* | | | | |
| Risk vectors | Vector 1 | Vector 2 | Vector 3 | Vector 4 |
| Nederland—Smart industry; or factories of the future 4.0 [104] | Not covered | Not covered | Not covered | SWAT analysis |
| Belgium—Made different [101] | Not covered | CPS | Not covered | Not covered |
| Spain—Industrie Conectada 4.0 [92] | Not covered | CPS | Linking the physical to the virtual to create intelligent industry | HADA—advanced self-diagnosis tool |
| Italy—Fabbrica Intelligente [96] | Not covered | Not covered | Not covered | Not covered |
| *IoT cyber risk vectors in elusive and roughly defined high-tech strategies* | | | | |
| Risk vectors | Vector 1 | Vector 2 | Vector 3 | Vector 4 |
| China—Made in China 2025 [102] | Not covered | Not covered | Automated machine tools and robotics | Financial and fiscal state control |
| G20—New industrial revolution (NIR) [105] | Not covered | Not covered | Not covered | Not covered |

## 5 Empirical study of cyber security standards

A key part of understanding the risks and issues facing the IoT involves reflecting on the standards and models present today. In what follows, we reflect on seven cyber risk standards and two cyber risk models. The design initiates with integrating best practices. Through empirical analysis [2], we compare existing cyber security measures and standards (e.g. FAIR and NIST cyber security frameworks) to propose a new and improved design principles for calculating the economic impact of IoT cyber risk.

The analysis presented in this section emerge from the analysis in this study, but also represent a stand-alone piece of work because the nature of security frameworks and assessment tools is quite diverse. What is presented in this section is an attempt to apply comparative analysis to synthesize a common, best practice approach that pulls the best features from each of the frameworks into a single, theoretical approach.

### 5.1 Empirical analysis of cyber security frameworks, models and methodologies

A majority of the cyber security frameworks today apply qualitative approaches to measuring cyber risk [114–118]. Some of the frameworks propose diverse qualitative methods, such as OCTAVE, which stands for Operationally Critical Threat, Asset, and Vulnerability Evaluation [115] and recommends three levels of risk (low, medium, high). Methodologies, such as the Threat Assessment and Remediation Analysis (TARA) [116] are also qualitative and apply a standardised template to record system threats. There also systems that combine qualitative and quantitative approaches. The Common Vulnerability Scoring System (CVSS) [119] provides modified base metrics for assigning metric values to real vulnerabilities. The CVSS applies expert's opinions, presented as statements, where each statement is allocated a level of cyber risk and the calculator assesses the overall level of risk form all statements.

Considering the lack of more precise methods, the modified base metrics represent the state of the art at present. The supply chain cyber risks are also assessed with qualitative approaches [39, 46, 120–127]. The Exostar system [128], which represent a qualitative approach, provides guidance points for assessing the supply chain cyber risk. The overall current state of cyber maturity can be verified with the Capability Maturity Model Integrated (CMMI) [118], which integrates five levels of the original Capability Maturity Model (CMM) [129]. To reach the required cyber security maturity level, the current cyber state can be transformed into a given a target cyber state by applying the National Institute of Standards and Technology's (NIST) [114] cyber security framework implementation guidance [130]. The risk assessment approach is based on the framework for improving cybersecurity of critical infrastructure [131] and follows recommendations for qualitative risk assessments e.g. standards based approach, or internal controls approach.

Slightly different approach to understanding risk is the use of emerging quantitative cyber risk models, such as the Factor Analysis of Information Risk Institute (FAIR) approach [132]. In effect, quantitative approaches are mostly present in the cyber security models [133, 134]. The FAIR approach is complementary to existing risk frameworks that are deliberately distanced from quantitative modelling (e.g. NIST) and applies knowledge from existing quantitative models, e.g. RiskLens [133], and Cyber VaR (CyVaR) [134]. In a way, FAIR is complementing the work of NIST and the International Organisation for Standardisation (ISO) [135], which is the international standard-setting body and includes cyber risk standards. For example, the ISO 27032 is a framework for collaboration that provides specific recommendations for cyber security, and ISO 27001 sets requirements for organisations to establish an Information Security Management System (ISMS).

Notable for this discussion, only ISO 27031 and NIST [114] provide recommendations for recovery planning, which some of the other frameworks and models have focused on less. A key point to note here is that risk estimation is used for recovery planning, and as such quantitative risk impact estimation [136] is needed for making decisions on topics such as cyber risk insurance [137]. The quantitative risk assessment approaches e.g. FAIR [132], RiskLens [133], and CyVaR [134], are still be in their infancy. Hence, the state of the art in current risk estimation (also known as risk analysis) is based on the high, medium, low scales (also known as the traffic lights system or colour system).

The diversity of approaches for cyber risk impact assessment, reemphasises the requirement for standardisation of cyber risk assessment approaches. The diversity and the gaps in the proposed approaches, become clearly visible in Table 2. This diversity presents conflict in risk assessment, e.g. qualitative versus quantitative. To enable the standardisation design, in Table 2, core cyber impact assessment concepts are extracted to defining the design principles for cyber risk impact assessment from IoT vectors. The design principles initiate with defining how to measure, standardise and compute cyber risk and how to recover. These are defined as:

**Table 2** Empirical analysis of cyber risk frameworks, methodologies, systems and models

| Frameworks | ISO | NIST | FAIR |
|---|---|---|---|
| Measure | ISO 27032 | Categorising | Financial |
| Standardise | ISO 27001 | Assembling | Complementary |
| Compute | Compliance | Compliance | Quantitative |
| Recover | ISO 27031 | Compliance | Level of exposure |
| Methodologies | TARA | CMMI | OCTAVE |
| Measure | Threat matrix | Maturity models | Workshops |
| Standardise | Template threats | ISO 15504—SPICE | Repeatability |
| Compute | Qualitative | Maturity levels | Qualitative |
| Recover | System recovery | Refers to other standards | Impact areas |
| Systems | Exostar system | CVSS calculator | |
| Measure | ISO 27032 | Base metrics | |
| Standardise | ISO 27001 | Mathematical approximation | |
| Compute | Compliance | Qualitative | |
| Recover | ISO 27031 | Not included | |
| Models | RiskLens | CyVaR | |
| Measure | BetaPERT distributions | VaR | |
| Standardise | Adopt FAIR | World economic forum | |
| Compute | Quantitative risk analytics with Monte Carlo and sensitivity analysis | Quantitative risk analytics with Monte Carlo | |
| Recover | Not included | Not included | |

- Measure—calculate economic impact of cyber risk.
- Standardise—international cyber risk assessment approach.
- Compute—quantify cyber risk.
- Recover—plan for impact of cyber-attacks, e.g. cyber insurance.

Beyond these issues, the empirical research outlined in Table 2 has highlighted other challenges in adopting existing cyber risk frameworks for dynamic and connected systems, where the IoT presents great complexities. For example the increasing ability of risk to propogate given the high degrees of connectivity in digital, cyber-physical, and social systems, and challenges pertaining to the limited knowledge that risk assessors have of dynamic IoT systems [13].

## 6 Theoretical analysis to uncover the best method to define a unified cyber risk assessment

The above empirical and comparative analysis correlated academic literature with government and industry cyber security frameworks, models and methodologies. In this section, epistemological analysis is applied to probe the existing understanding of cyber risk assessment. Such an approach was considered appropriate for our purposes because most cyber security frameworks and methodologies propose answers to a quantitative question with qualitative assessments. The analysis in this study examines how the current cyber risk assessment approaches are based on conventional abstractions, for instance, the colour coding in the NIST framework traffic light protocol [138], or the mathematical approximation in CVSS [119]. In quantified cases, we may have a modified attack vector allocated to a numerical value of 0.85 for a network metric value, and a numerical value of 0.62 for adjacent network metric value [117]. The question is why 0.85 and why 0.62 and why red represents information not for disclosure [138]. These measurements represent conventional abstractions that when expressed, become important units of measurement. These units of measurement in effect represent symbols with a defined set of rules in a conventional system, where truths about their validity can be derived from expert opinions, hence proven to be correct. These units of measurement do not, however, represent quantitative units based on statistical methods for predicting uncertainty.

Knowledge requires 'truth, belief and justification' as individual conditions [90]. Knowledge that a numerical value of 0.62 is 'true' metric value for adjacent network, as the related CVSS approach 'believes', needs to be 'justified' to confirm it does not represent just a guess of luck. Since a numerical value. Justification needs to be based on evidentialism [139, 140], where a proposition e.g. numerical value of 0.62, is epistemically justified as determined entirely by evidence. The debate whether cyber risk standards can be epistemically justified, must be based on the facts and evidence currently available. In evidentialism, epistemic evaluations are separate from moral believes and practical assessments, as epistemically justified evaluations might conflict with moral and practical estimations [139].

## 7 Epistemological framework

The integration of the theoretical analysis, with the empirical study of existing models with the comparative study of national strategies leads to a new epistemological framework consistent of sets of techniques for impact assessment of IoT cyber risk. Subsequently, a grounded theory approach is applied on the results of the epistemological framework with the output of the case study research into IoT cyber trends and technologies. The case study research is not applied to identify new, or the most prominent risk vectors. It would be challenging to argue that there is no bias if the vectors came from a limited population of stakeholders. The case study research simply represents an example of how the epistemological framework could be applied in a step by step process.

### 7.1 Proposed epistemological framework for cyber risk assessment standardisation

To define a standardisation framework, firstly the Pugh controlled convergence [141] is applied with a group of experts in the field. The Pugh controlled convergence is a time-tested method for concept selection and for validation of research design. The results from the comparative study and the empirical analysis were presented, including the Fig. 1, to a group of experts. The Pugh controlled convergence [141] was applied to organise the emerging concepts into definitions of the design principles. The resulting definition of design principles for a standardisation framework are derived from four workshops that included 18 distinguished engineers from Cisco Systems, and 2 distinguished engineers from Fujitsu. The workshops with Cisco Systems were conducted in the USA in four different Cisco research centres. The Fujitsu workshop was conducted separately to avoid those experts being influenced or outspoken by the larger group from Cisco systems.

This approach to pursuing validity follows existing literature on this topics [40, 142] and provides clear definitions that specify the units of analysis for IoT cyber risk vectors. The reason for pursuing clarity on the units of analysis for IoT cyber risk, was justified by existing literature, where these are identified as recommended areas for further research [143]. The IoT risk units of analysis from individual high-tech strategy are combined into standardisation vectors. The process of defining the standardisation vectors followed the Pugh controlled convergence method, where experts were asked to confirm the valid concept, merge duplicated concepts, and delete conflicting concepts.

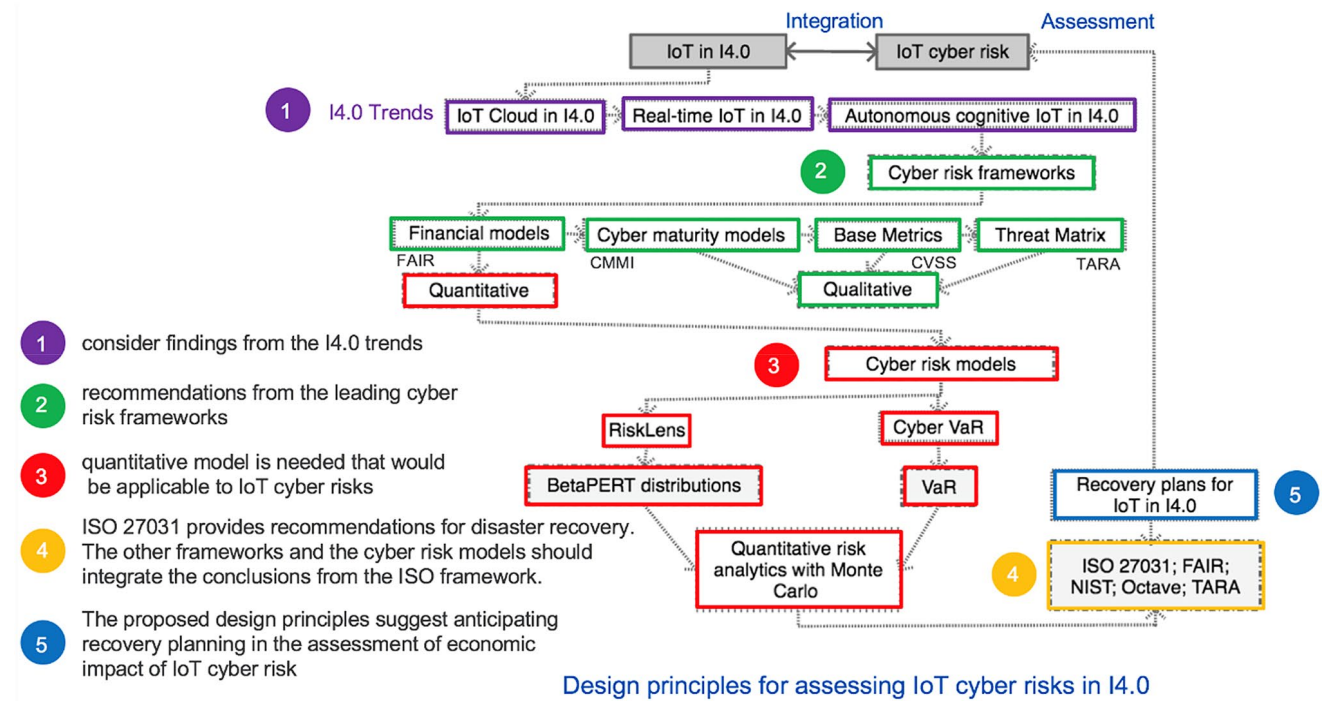

**Fig. 1** Design principles for assessing IoT cyber risks vectors in national strategies

In the assessment and transcription process, discourse analysis is applied to interpret the data [144] and for recognising the most profound concepts in the data [145]. The findings from the workshops are summarised in the table below (Table 3). The findings are presented in Table 3 after the controlled convergence was performed on all five workshops. The controlled convergence resulted with some units of analysis being merged to avoid duplication, such as Cloud-based computing [108]; and Cloud computing [86]. Or the concepts of CPS, which was identified as vector 2 in multiple high-tech strategies [86, 92, 103]. Similarly, the units of analysis of cyber risk standards are presented as merged definitions of the design principles, as categorised on the controlled convergence workshops.

The Table 3 below presents an epistemological framework of the knowledge and understanding, gathered from the comparative empirical analysis. The epistemological framework in Table 3 presents a narrowed framework of current understanding of IoT cyber risk, which is analysed and verified with the Pugh controlled convergence method for concept selection and for validation of research design.

The epistemological framework in Table 3 defines the IoT cyber risk vectors and relates the risk vectors with units of analysis. Defining the IoT cyber risk vectors and the related units of analysis, represents a crucial milestone in defining the design principles for cyber risk assessment of IoT. The epistemological framework in Table 3 proposes the design principles for measuring, standardising, computing and recovering from IoT risk. An example of how the epistemological framework in Table 3 can be applied:

- Measure the 'vector 3': economic impact of cyber risk from autonomous 'robotics in IoT'—calculate economic impact of cyber risk with 'BetaPERT distributions'.
- Standardise the 'vector 3'—international cyber risk impact from autonomous 'robotics in IoT'—assessment approach with 'Mathematical approximation'.
- Compute the impact from 'vector 3': economic impact of cyber risk from autonomous 'robotics in IoT'—quantify cyber risk with 'Quantitative risk analytics with Monte Carlo and sensitivity analysis'.
- Recovery planning for the 'vector 3': calculate financial cost from cyber risk from autonomous 'robotics in IoT' and determine maximum acceptable 'level of exposure' for 'system recovery'—plan for cyber insurance for the determined 'level of exposure'.

Table 3 Epistemological framework for standardisation of cyber risk impact assessment

| IoT cyber risk | | | | |
|---|---|---|---|---|
| Cyber risk vectors | Vector 1 | Vector 2 | Vector 3 | Vector 4 |
| | Cloud | Real-time | Autonomous | Recovery |
| Vector units of analysis | Cloud-computing platforms; technology skills; data centres; software; guidance; monitoring; Integration in cloud computing; Society 5.0; security networks | Operational models in real time; Customised products in real time; Digital real-time and interoperable records; Platform for real-time information; Connected industries; CPS | Automated environments; Robotics and Autonomous Systems; Robotics and artificial intelligence; Active cyber defence; Robots innovation; Robot society; Robotics in IoT; Artificial intelligence and control systems | Economic impact; Impact assessment; SWAT analysis; HADA—Advanced self-diagnosis tool; Financial and fiscal state control |
| *Standardisation framework for cyber risk assessment* | | | | |
| Measure | ISO 27032; Categorising; financial; threat matrix; maturity models; workshops; ISO 27032; base metrics; BetaPERT distributions; VaR | | | |
| Standardise | ISO 27001; assembling; complementary; template threats; ISO 15504—SPICE; repeatability; ISO 27001; mathematical approximation; adopt FAIR; world economic forum | | | |
| Compute | Compliance; quantitative; maturity levels; qualitative; quantitative risk analytics with Monte Carlo and sensitivity analysis | | | |
| Recover | ISO 27031; compliance; level of exposure; system recovery; impact areas | | | |

This example covers only one risk vector and one unit of analysis. The example is appropriate for an enterprise that aims to deploy autonomous robotics in IoT. National high-tech strategies would need to perform all analysis, for all risk vectors, with all units of analysis provided in the epistemological framework in Table 3. It is surprising that national high-tech strategies have not until present performed such analysis. Especially concerning are the findings from the gap analysis in Table 1 which confirms that many of the areas covered by the epistemological framework in Table 3 are not covered in some of the national high-tech national strategies. An example of how such analysis could be performed in provided in Fig. 1 below. This design process follows recommendations from literature [146], and shows how individual cyber risk components can be integrated into an impact assessment standardisation infrastructure. The epistemological framework is promoting the development of a generally accepted cyber security approach. This is also called for in current research work [11–13], because the IoT adoption requires standardisation reference architecture [53, 86, 147, 148] to encompass security and privacy [69].

### 7.2 Defining the design principles for cyber risk assessment of IoT vectors

In the section above, we propose a new set of design principles for assessing the cyber risk from IoT risk vectors. The principles had been tested through workshops and a comparative study to ensure the process can be applied in real-world practice. The comparative study shows that IoT trends have failed to implement the recovery planning. This is in contradiction with the findings from the second reflection of the empirical study of cyber risk assessment standards, where the recovery planning is strongly emphasised (*see:* ISO, FAIR, NIST, OCTAVE, TARA). It seems that the IoT high-tech strategies may have overlooked the recommendations from the cyber risk assessment standards. A standardisation approach for IoT impact assessment should firstly consider the new IoT cyber risk vectors derived from the comparative study. Secondly, a standardisation approach should consider the recommendations from the empirical study. The empirical study recommends a decomposition process of assessment standards, conducting grounded theory analysis. This was followed by a compounding of concepts to address individual gaps in cyber risk assessment standards.

The empirical and comparative study investigated the soundness of current cyber risk assessments. The theoretical analysis however, was applied to probe the soundness of the qualitative versus quantitative assessment approaches. Theoretical analysis confirmed that to identify the cost of recovery planning and/or the cost of cyber insurance, a new quantitative model is needed that would anticipate IoT risks. The empirical study analysed the leading quantitative cyber risk models (RiskLens, supported by FAIR; and CyVaR, supported by the World Economic Forum, Deloitte and FAIR). The unifying link between the two cyber risk models was identified as the application of Monte Carlo simulations, for predicting cyber risk uncertainty. A new impact assessment model for the IoT risk vectors, should implement the guidance from RiskLense and CyVaR. The main guidance is that the application of Monte Carlo simulation would reduce the IoT cyber risk uncertainty and enable the approximation and estimation of the economic impact of cyber risk from IoT devices. Such calculation would enable companies to develop appropriate recovery planning and the insurance industry to provide a more realistic cost of cyber insurance.

At a higher analytical level, in Fig. 1 we propose a new set of design principles for assessing the cyber risk from IoT risk vectors. The comparative study of IoT in national high-tech strategies shows that I4.0 trends have failed to implement the recovery planning in the leading national initiatives. This is in contradiction with the findings from the second reflection from the empirical study of the leading cyber risk frameworks, where the recovery planning is strongly emphasised (ISO, FAIR, NIST, Octave, TARA). It seems that the leading high-tech strategies initiatives have ignored the recommendations from the world leading cyber risk frameworks. A new model for IoT risk impact assessment should firstly consider the findings from the comparative study of I4.0 trends, secondly the recommendations from the empirical study of leading cyber risk frameworks. To identify the cost of recovery planning or the cost of cyber insurance, a new quantitative model is needed that would be applicable to IoT cyber risks. There are currently two leading quantitative cyber risk models. First is the RiskLens approach, promoted by FAIR. Second is the Cyber VaR, promoted by the World Economic Forum, Deloitte and more recently by FAIR. The unifying link between the two cyber risk models is the application of Monte Carlo simulations for predicting cyber risk uncertainty. From this study, it appears that a new impact assessment model for the cyber risks from IoT in high-tech national strategies, should start with the guidance from RiskLense and Cyber VaR. The application of Monte Carlo simulation would reduce the IoT cyber risk uncertainty and enable the approximation and estimation of the economic impact of cyber risk from IoT devices. Such calculation would enable companies to develop appropriate recovery planning and the insurance industry to provide a more realistic cost of cyber insurance.

The proposed design principles suggest anticipating recovery planning in the assessment of economic impact of IoT cyber risk. Such approach would enable cyber insurance companies to value the impact of IoT cyber risks in I4.0. The rationale of the proposed design principles is that without appropriate recovery planning, the economic impact can be miscalculated, resulting in greater losses than we anticipated initially. The design principles are developed to advance the existing efforts (from the World Economic Forum, Deloitte, FAIR, etc.) in developing a standardised quantitative approach for assessing the impact of cyber risks. The described design process decomposes the most prominent risk vectors and units of analysis and formulates a generalised set of IoT risk vectors. This does not refer to a complete set of vectors, but to the most prominent risk vectors. Considering that such study has not been conducted until present, the process of integrating the most prominent vectors, with a comparative analysis of the a diverse set of security frameworks and tools, represents the first step in understanding the standardisation process. The design principles in Fig. 1, also present multiple approaches to calculating the economic risk of IoT implementation (e.g. BetaPert, Cyber VaR, RiskLense). This connects the design with the described issues related to the costs of risk.

### 7.3 Use cases

Case study research is applied for extending, evaluating and comparing the framework with the other frameworks listed in the empirical analysis (Sect. 5.1—Table 2). The industrial case study requesting the participants to apply the framework to their cyber risk assessment of IoT risk. To clarify the participants understanding of the framework, a series of open-ended interviews were performed. The pool of participants interviewed were proportionally representative of different levels of seniority. The initial participants were selected through convenience sampling. Only part of the interviews were predetermined in the initial selection and the rest were chosen based on the development of the case study research. The industrial case study involved series of 20 qualitative interviews, followed by 4 group discussions, two with experts from Cisco Systems in the USA; one with experts from Fujitsu centre for Artificial Intelligence the UK and second with Fujitsu Coelition (I4.0 centre) in the UK. The data collected was transcribed and categorised with aims to investigate the relationship between the notion of IoT and existing cyber risk assessments with their company established approaches (see Table 2 for all approaches investigated and compared in the case study). The aim of the analysis was to verify the ideas behind the epistemological framework and to relate IoT technologies to established models for cyber risk assessment.

The generic diagram from Fig. 1 was presented to the participants and it enables quick comparative analysis of the entire epistemological design process. This enabled practitioners to compare the new framework, with the established cyber risk assessment approaches (in Table 2). The design process for IoT risk assessment (in Fig. 1) is generic and could be applied by other companies and sectors. The generic design outlines a new approach for cyber risk assessment from the IoT. This was considered by the participants as easier to understand and navigate through than the cyber risk assessment approaches (from Table 2). The main feedback from the use cases was that this framework enables easier understanding of the specifics of IoT cyber risk with the direct reference points to the required type of assessment. While the existing cyber risk assessment approaches that were compared (see Table 2) were considered more time demanding. While many comprehensive frameworks and modes are currently in use, from our use case study we determined that many experts do not understand the technicalities and the continuous updates of the established frameworks. The epistemological framework we presented on the use case, was confirmed as easier, quicker, and more practical for the participating practitioners.

### 7.4 Discussion

The research problem investigated in this article was the present lack of standardised methodology that would measure the cost and probabilities of cyber-attacks in specific IoT related verticals (ex. connected spaces or commercial and industrial IoT equipment) and the economic impact (IoT product, service or platform related) of such cyber risk.

The lack of recovery planning is consistent in all of the high-tech strategies reviewed. Adding to this, the new risks emerging from IoT connected devices and services, and the lack of economic impact assessments from IoT cyber risks, makes it imperative to emphasise the lack of recovery planning. The volume of data generated by the IoT devices creates diverse challenges in variety of verticals (e.g. machine learning, ethics, business models). Simultaneously, to design and build cyber security architecture for complex coupled IoT systems, while understanding the economic impact, demands bold new solutions for optimisation and decision making [13]. Much of the research is application-oriented and by default interdisciplinary, requiring hybrid research in different academic areas. Hybrid and interdisciplinary approaches are required, for the design of cyber risk assessment that integrate economic impact from IoT verticals. Such design must meet public acceptability, security standards, and legal scrutiny.

As a result of the fast growth of the IoT, cyber risk finance and insurance markets are lacking empirical data to construct actuarial tables. Despite the development of models related to the impact of cyber risk, there is a lack of such models related to specific IoT verticals. Hence, banks and insurers are unable to price IoT cyber risk with the same precision as in traditional insurance lines. Even more concerning, the current macroeconomic costs estimates of cyber-attacks related to IoT products, services and platforms are entirely speculative. The approach by 'early adopters' that IoT products are 'secure by default' is misleading. Even governments advocate security standards ex. standards like ISA 99, or C2M2 [129, 149] that accept that the truth on the ground is that IoT devices are unable to secure themselves, so the logical placement of security capability is in the communications network.

## 8 Conclusion

This article decomposes the cyber risk assessment standards and combines concepts for the purposes of building a model for the standardisation of impact assessment approaches. The model enables the implementation of two current problems with assessing cyber risk from IoT devices. First, the model enables identifying and capturing the IoT cyber risk from different risk vectors. Second, the model offers new design principles for assessing cyber risk. The analysis in this paper was focused on understanding the best approach for quantifying the impact of cyber risk in the IoT space. The model and the documented process represents a new design for mapping IoT risk vectors and optimising IoT risk impact assessment.

The model proposes a process for adapting existing cyber security practices and standards to include IoT cyber risk. Despite the interest to standardise existing cyber risk frameworks, models and methodologies, this has not been done until present. Standardisation framework currently does not exist in literature and the epistemological framework represents the first attempt to define the standardisation process for cyber risk impact assessment of IoT vectors. This article applies empirical (via literature reviews and workshops) and comparative studies with theoretical analysis and the grounded theory to define a process of standardisation of cyber risk impact assessment approaches. The study advances the efforts of integrating standards and governance on IoT cyber risk and offers a better understanding of a holistic impact assessment approach for cyber risk. The documented process represents a new design for mapping and optimising IoT cyber security.

The empirical study defined the gaps in current cyber risk assessment frameworks, models and methodologies. The identified gaps are analysed to define a process of decomposing risks and compounding assessment concepts, to address the gaps in cyber risk standards. The comparative study defines the IoT cyber risk vectors which are not anticipated or considered in existing cyber risk assessment standards. The epistemological analysis adapts the current cyber security standards and defines the principles for integrating specific IoT risk vectors in a holistic impact assessment framework. It is anticipated that the analysis of the complete economic impact of data compromise of IoT risk vectors, would empower the communications network providers to create clear, rigorous, industry-accepted mechanisms to measure, control, analyse, distribute and manage critical data needed to develop, deploy and operate cost-effective cyber security for critical infrastructure. The research design identifies and captures the IoT cyber risk vectors and defines a framework for adapting existing cyber risk standards to include IoT cyber risk.

### 8.1 Limitations and further research

The epistemological framework in this article is derived from empirical and comparative studies, supported with theoretical analysis of a limited set of frameworks, models, methodologies and high-tech strategies. The set selection was based on documented availability and on relevance to cyber risk impact assessment of IoT risk vectors. Holistic analysis of all risk assessment approaches was considered beyond the scope of this study. Additional research is required to integrate the knowledge from other studies.

**Acknowledgements** Sincere gratitude to the Fulbright Project.

**Funding** This work was supported by the UK EPSRC [project grant number: EP/S035362/1, EP/N023013/1, EP/N02334X/1] and by the Cisco Research Centre [Project grant number: 1525381]. Earlier versions of this work from the combined working papers and project reports prepared for the PETRAS National Centre of Excellence and the Cisco Research Centre can be found in a pre-print online [38].

## Compliance with ethical standards